\documentclass[journal]{IEEEtran}

\usepackage{amsmath}
\usepackage{graphicx}
\usepackage{subfigure}
\usepackage{algpseudocode}
\usepackage{algorithmicx,algorithm}
\usepackage{amssymb}
\usepackage{float}
 \usepackage{verbatim}
\usepackage{multirow}
\usepackage{makecell}
\usepackage{hyperref}
\usepackage{graphics}
\usepackage{diagbox}
\usepackage{color}

\usepackage{cleveref}

\usepackage[numbers,sort&compress]{natbib}
\begin{document}
\title{Deep Learning-Based Frequency Offset Estimation}

\author{Tao~Chen,
        Shilian~Zheng,
        Jiawei~Zhu,
        Qi~Xuan,
        and Xiaoniu~Yang
\thanks{This work was supported in part by the National Natural Science Foundation of China under Grants U20B2038 and  U21B2001 and by the Key R\&D Program of Zhejiang under Grant 2022C01018. \textit{(Corresponding author: Shilian Zheng.)}}
\thanks{T. Chen and Q. Xuan are with the Institute of Cyberspace Security, and also with the College of Information Engineering, Zhejiang University of Technology, Hangzhou 310023, China (e-mail: ctchentao369@163.com; xuanqi@zjut.edu.cn).}
\thanks{S. Zheng, J. Zhu, and X. Yang are with Innovation Studio of Academician Yang, National Key Laboratory of Electromagnetic Space Security,
Jiaxing 314033, China (e-mail: lianshizheng@126.com;  zhujiaweigl@126.com; yxn2117@126.com).}
}


\maketitle

\begin{abstract}
In wireless communication systems, the asynchronization of the oscillators in the transmitter and the receiver along with the Doppler shift due to relative movement may lead to the presence of carrier frequency offset (CFO) in the received signals. Estimation of CFO is crucial for subsequent processing such as coherent demodulation.
In this brief, we demonstrate the utilization of deep learning for CFO estimation by employing a residual network (ResNet) to learn and extract signal features from the raw in-phase (I) and quadrature (Q) components of the signals. We use multiple modulation schemes in the training set to make the trained model adaptable to multiple modulations or even new signals. In comparison to the commonly used traditional CFO estimation methods, our proposed IQ-ResNet method exhibits superior performance across various scenarios including different oversampling ratios, various signal lengths, and different channels. 

\end{abstract}


\section{Introduction}



\IEEEPARstart {I}{n} wireless communication systems, the oscillators used in the transmitter and receiver cannot achieve perfect synchronization due to real-world factors. This discrepancy can lead to a noticeable frequency deviation between the transmitter and the receiver \cite{2017Optimization}. Such carrier frequency offset (CFO) will lead to the rapid decline of coherent demodulation performance of the synchronous receiver, resulting in the unreliability of normal communication, and seriously reducing the performance of mobile communication systems. Therefore, it is very important to accurately estimate and compensate the CFO in communications systems.


CFO estimation technologies encompass two primary categories: data-aided (DA) \cite{2014A} and non-data-aided (NDA) \cite{2017A} techniques. The DA methods leverage known signal sequences to estimate the frequency offset, simplifying the operational and computational aspects while achieving enhanced estimation accuracy. Examples of DA methods include the maximum likelihood (ML) method \cite{1055282}, the autocorrelation method \cite{7784355}, and the Kay \cite{9231178} method. These approaches prove particularly suitable for carrier burst transmission systems. 
Nonetheless, they are notably vulnerable to noise and interference, leading to inadequate resistance against such disruptions \cite{8649676}. In non-cooperative communication systems, it is often not possible to acquire prior knowledge of the transmitted signal. In this case, NDA CFO estimation methods have gained widespread adoption, which relies solely on the received signal itself, eliminating the requirement for additional prior information. These methods allow for quicker and more efficient frequency offset estimation. For example, the delay multiplication structure is utilized for CFO estimation and has good performance in binary continuous phase modulation (CMP) signals \cite{8359649}. In addition, when the modulation information of the received data is eliminated \cite{2012Baseband}, some DA methods can be used to conduct CFO estimation in scenarios without data assistance. NDA Kay method \cite{937215} is one of these methods.
Similarly, these NDA CFO estimation methods yield unsatisfactory results when confronted with low SNR and exhibit limited adaptability to distortions due to fading environment.

With the rapid development of 
deep  learning (DL), it has been widely used in various fields such as natural language processing (NLP) \cite{9245375} and image recognition \cite{2015Delving}. DL mimics the intricate neural architecture of the human brain, enabling it to process vast volumes of data and extract high-level features. These acquired high-level representations are then utilized to undertake classification or regression tasks. The realm of communication has increasingly embraced DL, harnessing neural networks for tasks like identifying signal modulation \cite{9935275}, detecting anomalies \cite{6415763}, and beyond. Correspondingly, in the CFO estimation, DL has also been explored. For example, the authors in \cite{8580824} designed a CP-OFDM receiver using deep neural networks (DNN) to estimate CFO of received orthogonal frequency-division multiplexing (OFDM) signals with cyclic prefix (CP). Moreover, the fusion of convolutional neural network (CNN) and attention mechanism gave rise to the CAD model structure, which was utilized for estimating CFO values of OFDM signals \cite{9992153}.

 \begin{figure*}[t]
    \centering
    \includegraphics[width=17cm]{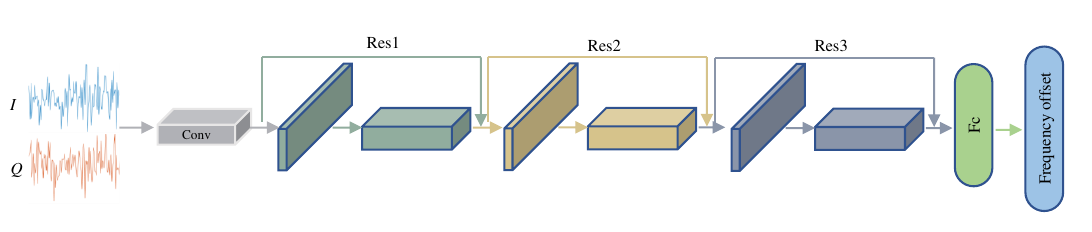}
   \caption{The overall framework of CFO estimation, where the convolution layer is denoted as Conv, the fully connected layer is denoted as Fc, the residual blocks are represented by Res1, Res2, and Res3, which refer to the first, second, and third residual blocks, respectively.}
    \label{fig1}
\end{figure*}
In the context of CFO estimation, while deep learning has been employed, the existing methods primarily concentrate on estimating the CFO of specific signals such as OFDM rather than offering a universally applicable CFO estimation method. 
In this brief, we introduce a simple deep learning-based method for estimating the CFO universally for multiple modulations in communication system. The method basically involves extracting the in-phase (I) and quadrature (Q) components of the signal and feeding these components into a ResNet network for estimation (denoted as IQ-ResNet method). In contrast to traditional CFO estimation methods, our proposed method exhibits superior performance in terms of estimation accuracy across various SNRs, different oversampling ratios, different signal lengths, and fading channels. With the augmentation of modulation diversity in the training dataset, a gradual enhancement in estimation performance becomes evident. This enhancement highlights the adaptability of the IQ-ResNet method to new kind of signals. 

\section{Method}
\subsection{Problem Description}
In the process of transmitting signals in a communication system, the signals will encounter the influence of channel and noise. Considering single carrier burst transmission system, the discrete signal obtained by the receiver after filtering and sampling can be expressed as:
\begin{equation}
\label{eq1}
r(n)=s(n)*  h(n)e^{j(2 \pi \Delta f n+\theta)}+w(n), n=0,1,\dots,L-1,
\end{equation}
where $r(n)$ is the received signal, $s(n)$ is the transmitted baseband complex signal,  $h(n)$ is the channel response, $\Delta f$ is the carrier frequency offset, $\theta$ is the carrier phase offset, $w(n)$ is additive Gaussian white noise (AWGN), and $L$ is the length of the received signal $r(n)$. The convolution operation is denoted by $*$.


CFO estimation employs the received signal to estimate the frequency offset of the signal. this process can be expressed as determining the predicted value with the highest probability as the estimated value:
\begin{equation}
\label{estimation}
\widehat{\Delta f}={\arg \max}_{\Delta f} \operatorname{Pr}\{\Delta f \mid r(n)\},
\end{equation}
where $ \widehat{\Delta f}$ is the estimated CFO, $\operatorname{Pr}\{\cdot\}$ represents the calculation of probability.

\subsection{Proposed Method}
We use deep learning-based regression method for CFO estimation. Fig. \ref{fig1} depicts the overall framework illustrating the utilization of neural networks for estimating the frequency offset of received signal. The received signal is converted into a parallel IQ sequence with dimensions of $(2 , L)$. Subsequently, we employ the ResNet network to extract the features of the IQ sequence. Ultimately, the fully connected layer yields the predicted frequency offset values as the output. This method is referred to as the IQ-ResNet CFO estimation method in the rest of the brief.

For more convenient calculation and data processing, we extract the I and Q components of the received signals and stack them in parallel into $(2 , L)$ dimensional vectors. The operation of extracting IQ components of signals can be expressed as 
\begin{equation}
\label{IQ}
IQ(n)=\left[\begin{array}{c}
 I(n)\\
Q(n)
\end{array}\right]=\left[\begin{array}{c}
 \text{real}(r(n))\\
\text{imag}(r(n))
\end{array}\right],
\end{equation}
where $\text{real}(\cdot)$ and $\text{imag}(\cdot)$ denote the operations that extracting the real and imaginary parts of the signal, respectively. The extracted IQ components are formulated into a matrix which is used as the input of the network.

\begin{figure}[t]
\centering
\includegraphics[width=0.2\textwidth]{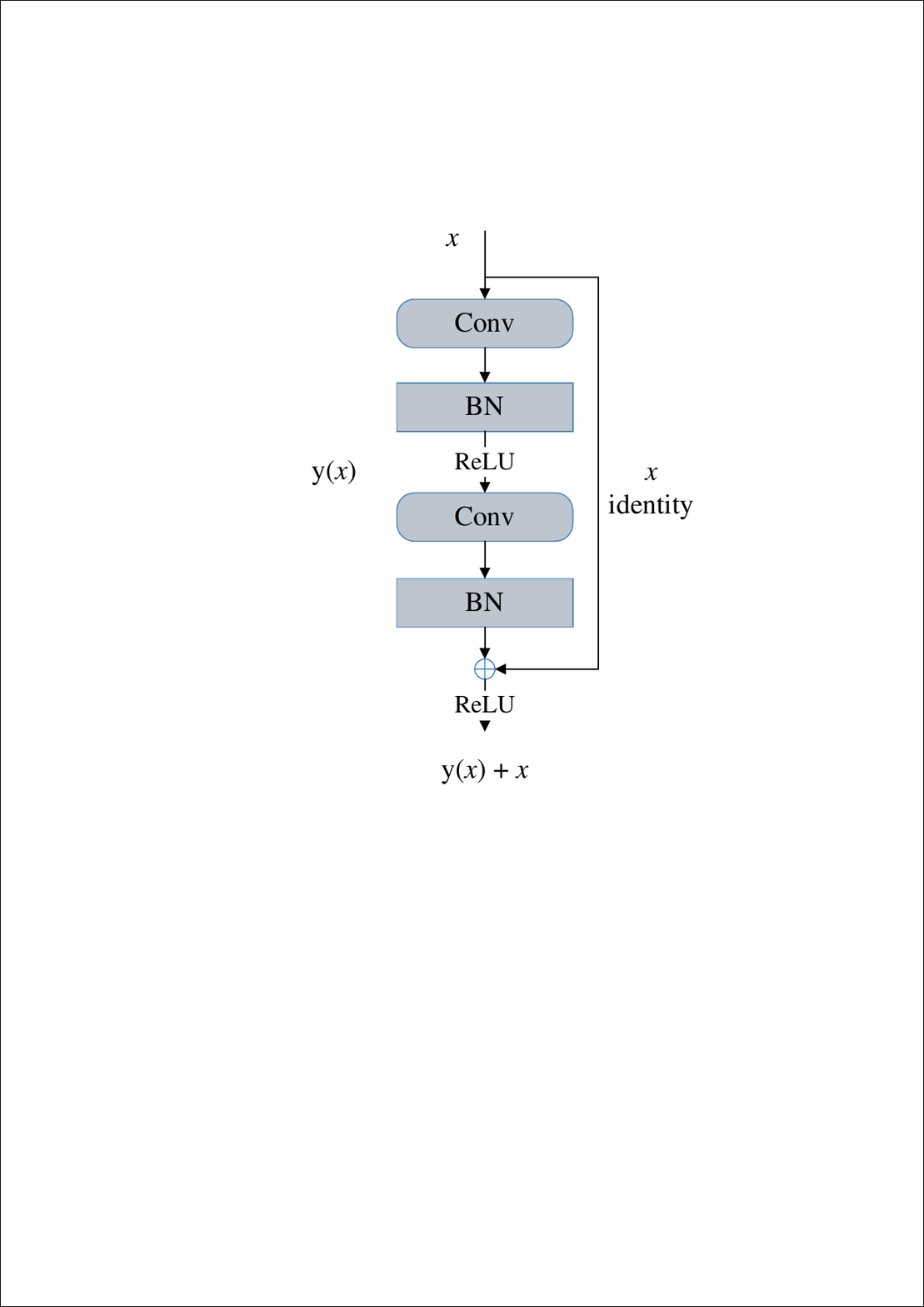}
\caption{\label{residual}The structure of the residual block.}
\end{figure}

ResNet is developed to address the problem of vanishing gradients in very deep CNNs. The main idea behind ResNet is the introduction of residual connections, also known as skip connections, that allow the network to learn residual functions. The key benefit of using residual connections is that they mitigate the degradation problem that occurs when networks become too deep. The structure of the residual blocks we used is shown in Fig \ref{residual}, which includes two groups of the convolution layer and batch normalization (BN) layer, which are connected by ReLU activation function. The ResNet we designed consists of three residual blocks, namely Res1, Res2, and Res3. The specific structure is shown in Table \ref{net}.

\begin{table}[t]
\renewcommand\arraystretch{2}
\centering
\caption{Specific Detail of ResNet Network}\
\label{net}
\setlength{\tabcolsep}{10mm}{
\begin{tabular}{cc}
\hline\hline
Layer & Output  \\ \hline
Input  &  1*2*$L$ \\ \hline
Conv   & 16*1*$L$  \\ \hline
Res1    & 16*1*$L$  \\ \hline
Res2    & 32*1*$L/2$  \\ \hline
Res1    & 64*1*$L/4$  \\ \hline
Fc    & 1  \\ \hline
\end{tabular}}
\end{table}

To ensure the receiver acquire the accurate carrier frequency, it is essential to estimate the precise CFO value using the received signal. To accomplish this, we employ regression method to forecast the carrier frequency offset values, which can be formulated as
\begin{equation}
\label{eq30}
 \widehat{\Delta f}=\mathcal{R}( IQ (n) ; \mathcal{O}),
\end{equation}
where $ \widehat{\Delta f}$ is the CFO estimation value predicted by ResNet, $\mathcal{O}$ is the parameter of the ResNet model, and $\mathcal{R}$ is the mapping of input and output in the ResNet. During the training phase, we employ the Mean Squared Error (MSE) as the loss function for the neural network. The expression of loss function is
\begin{equation}
\label{mse}
 \mathrm{loss}=\frac{1}{S} \sum_{i=1}^S\left(\Delta f_i-\widehat{\Delta f}_i\right)^2,
\end{equation}
where $S$ denotes the total number of samples fed into the ResNet network within each batch, $\widehat{\Delta f}_i$ is the predicted frequency offset value of the $i$-th sample, $\Delta f_i$ is the true frequency offset value of the $i$-th sample. This allows us to quantify the disparity between the predicted frequency offset and the actual frequency offset.

Training is the key to ensuring the performance of the IQ-ResNet model. According to the IQ components and the CFO labels, we can construct the training set for a specific modulation type as
\begin{equation}
\mathcal{D}_j=\left\{\left(IQ_{ij}, \Delta f_{ij}\right)\right\}_{i=1}^{N},
\end{equation}
where $IQ_{ij}$ is the $i$-th received signal in the $j$-th class of modulation type, $N$ is the number of samples in the dataset, and $\Delta f_{ij}$ is the true frequency offset value of $IQ_{ij}$. To improve the model's adaptability to different modulations and estimation accuracy, we embed samples of different modulation types to a single training dataset. Therefore, the expanding training dataset of different modulation types can be represented as
\begin{equation}
\mathcal{D}=\left\{\mathcal{D}_j\right\}_{j=1}^{M}=\left\{\left(IQ_{i1}, \Delta f_{i1}\right),\dots, \left(IQ_{iM}, \Delta f_{iM}\right)\right\}_{i=1}^{N}.
\end{equation}
where $M$ is the number of modulation types in the training samples. This expanding training dataset with signals of multiple modulation types is used to train the IQ-ResNet model. By doing this, the trained model can predict the CFO of the signal blindly no matter what modulation is adopted, thus offering a universal CFO estimation.

\section{Simulation}
\subsection{Simulation Setting}
To validate the effectiveness of our proposed IQ-ResNet method for predicting CFO values, we generate datasets similar as HKDD$\_$AMC12 \cite{10042021} dataset. 
We choose four modulations for experiments, i.e., BPSK, 2FSK, 16QAM and 4PAM. A root raised-cosine (RRC) filter with a truncated length of 6 symbols is utilized. The roll-off factor is randomly selected within a range of 0.2 to 0.7. The normalized frequency offset, ranging from $-$0.2 to 0.2, is randomly generated for each signal. The SNR varies from $-$20 dB to 30 dB, with a 2 dB interval between each SNR value. The training set consists of 312000 samples, while the test set comprises 156000 samples. We explore the impact of various over-sampling rates, signal length, as well as the presence of AWGN and Rayleigh channels in our experiment. The subsequent experimental analysis will provide the specific parameter settings for each corresponding dataset.


The simulation experiments are performed using an Intel I9-11900 CPU, Nvidia RTX3080Ti GPU, and 128GB of memory. During the initialization of the model, we configure the epoch value to 20 and set the initial learning rate to 0.02. Additionally, at the 5-th and 10-th epochs, the learning rate is reduced to 10\% of its previous value. We choose a minimum batch size of 64 and utilize the Adam optimizer.

\subsection{Simulation Results}
\subsubsection{Comparison with Traditional Methods}

\begin{figure}
\centering
\includegraphics[width=0.5\textwidth]{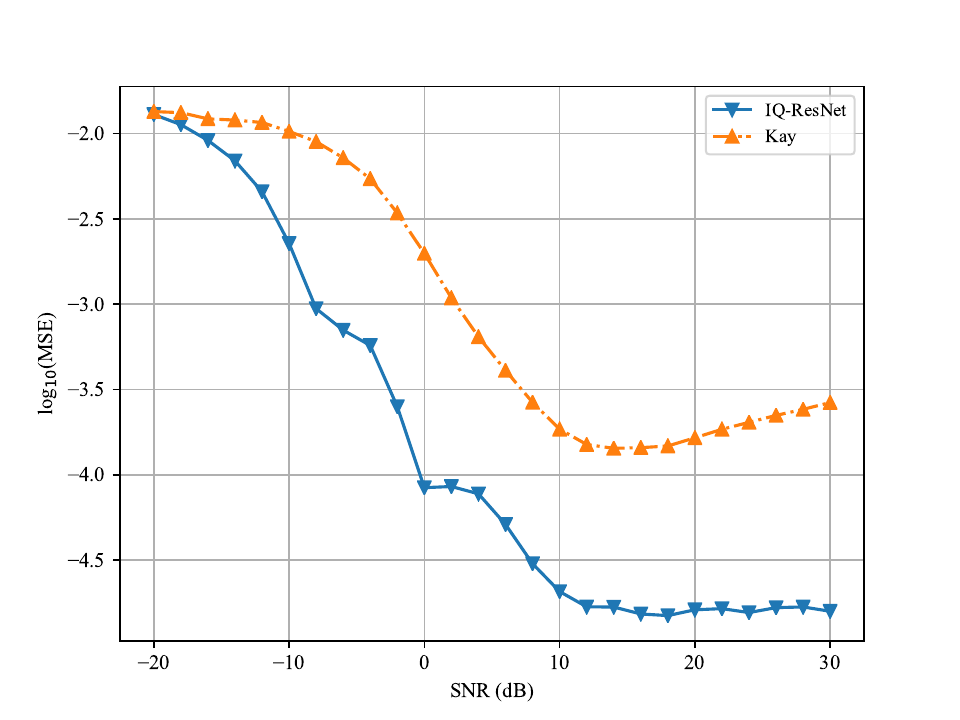}
\caption{\label{all}The MSE values of IQ-ResNet and Kay methods for all modulation types.}
\end{figure}

We first assess the effectiveness of our proposed IQ-ResNet method for CFO estimation by comparing its performance against traditional Kay-based method. We set $L$ to 1024 and $R$ to 8 and we assume an AWGN channel. We plot the MSE values for all modulation schemes using two different methods across different SNRs in Fig. \ref{all}. The figure clearly illustrates that the IQ-ResNet method outperforms the Kay-based method. Notably, as the SNR increases, the gap in performance between IQ-ResNet method and Kay-based method becomes remarkably pronounced. Meanwhile, the IQ-ResNet method exhibits remarkably low MSE values across the SNR range of 10 dB to 20 dB.





\begin{figure}
\centering
\subfigure[]{\label{RNSR_SNR:RNSR_RML1024}
\includegraphics[width=0.8\linewidth]{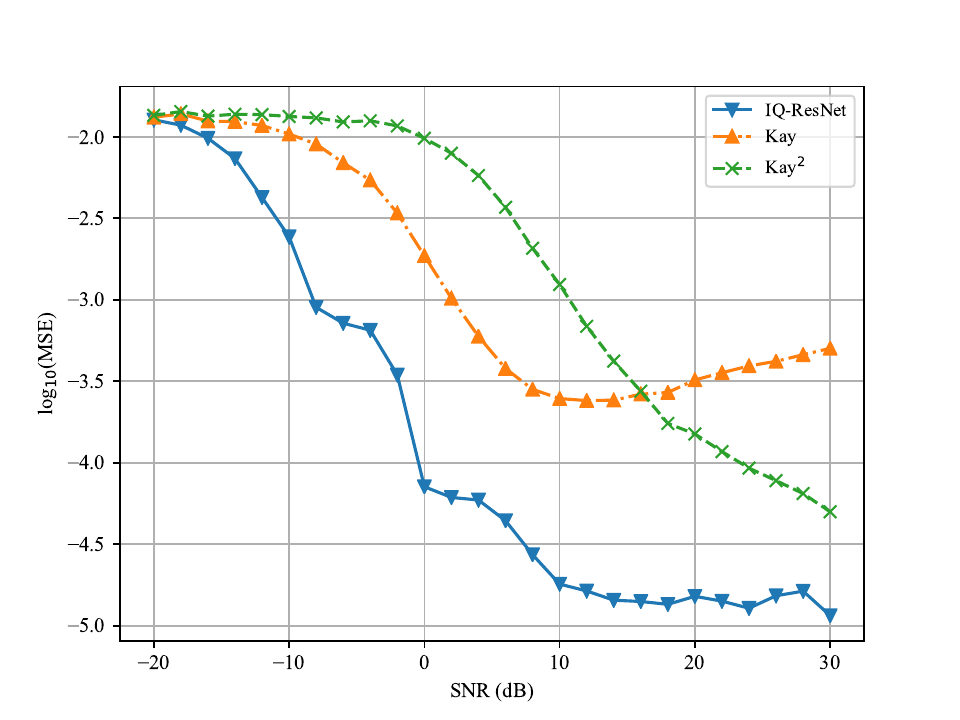}}
\subfigure[]{\label{fig1:subfig:RNSR_HKDD}
\includegraphics[width=0.8\linewidth]{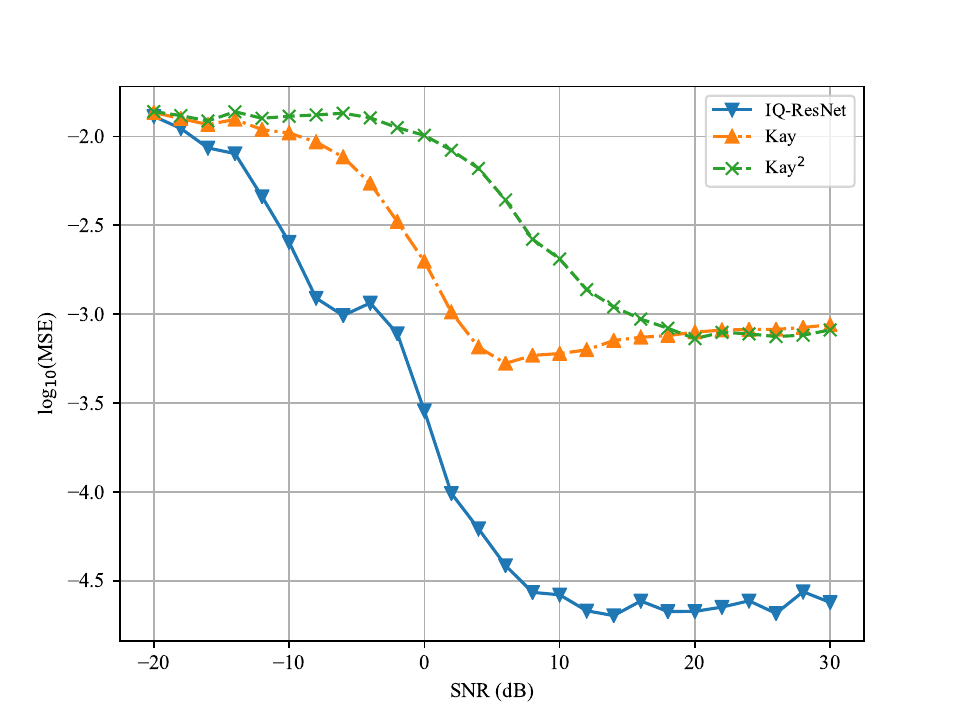}}
\centering

\caption{The MSE values of IQ-ResNet, Kay and Kay$^2$ methods for BPSK signals. (a) AWGN and (b) Rayleigh.}
\label{Specific}
\end{figure}

We next investigate the effectiveness of the CFO estimation method on a particular modulation, say BPSK. The Kay method after the square of BPSK signals (denoted as Kay$^2$-based method) is also provided for comparison. Both AWGN and Rayleigh channels are considered in this experiment. The results are shown in Fig. \ref{Specific}. It is clear that our IQ-ResNet method demonstrates superior performance compared to the other two estimation methods across the overall SNR. On AWGN channel, the Kay-based method outperforms the Kay$^2$-based method in low SNR but falls short compared to the Kay$^2$-based method in high SNR. Notably, the Kay-based method's performance declines as SNR increase. Conversely, the Kay$^2$-based method excels at high SNR and gradually converges toward the IQ-ResNet method's performance level. On the Rayleigh channel, the performance of the traditional methods decreaces significantly, while the performance of our proposed method is close to that of AWGN, further demonstrating the superiority of our method.



\subsubsection{Effect of Different Over-Sampling Rates}
\begin{figure}[t]
\centering
\includegraphics[width=0.4\textwidth]{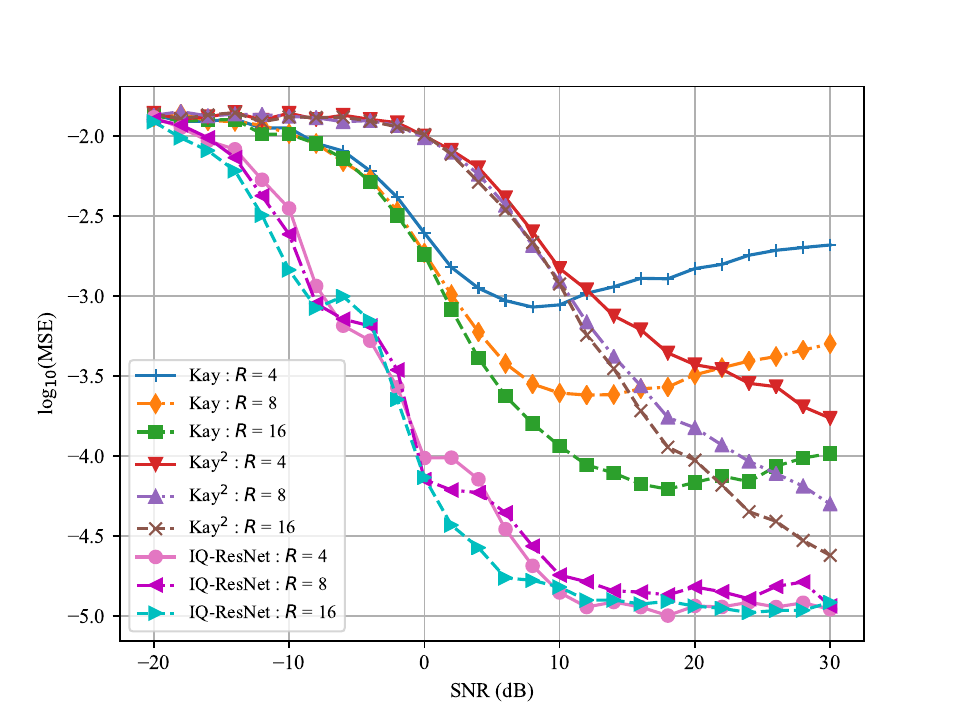}
\caption{\label{rate}Effect of different over-sampling rates.}
\end{figure}

We conduct experiments using over-sampling ratios of 4, 8, and 16 to investigate how different over-sampling rates affect the performance of CFO estimation. BPSK is used and the signal length is chosen to be 1024. The MSE values are shown in Fig. \ref{rate}. It is evident that varying over-sample rates minimally affect the IQ-ResNet method, preserving its commendable performance. On the contrary, the other two methods are significantly affected by changes in the over-sampling rate. Specifically, as the over-sampling rate increases, especially at high SNR, the MSE values for both the Kay-based method and Kay$^2$-based method gradually decrease. This effect is particularly pronounced in the Kay-based method. 
This result shows that IQ-ResNet method can still precisely extract features for CFO estimation at low over-sampling ratio, demonstrating its robustness and effectiveness for different over-sampling rates.

\subsubsection{Effect of Different Sample Lengths}
\begin{figure}[t]
\centering
\includegraphics[width=0.4\textwidth]{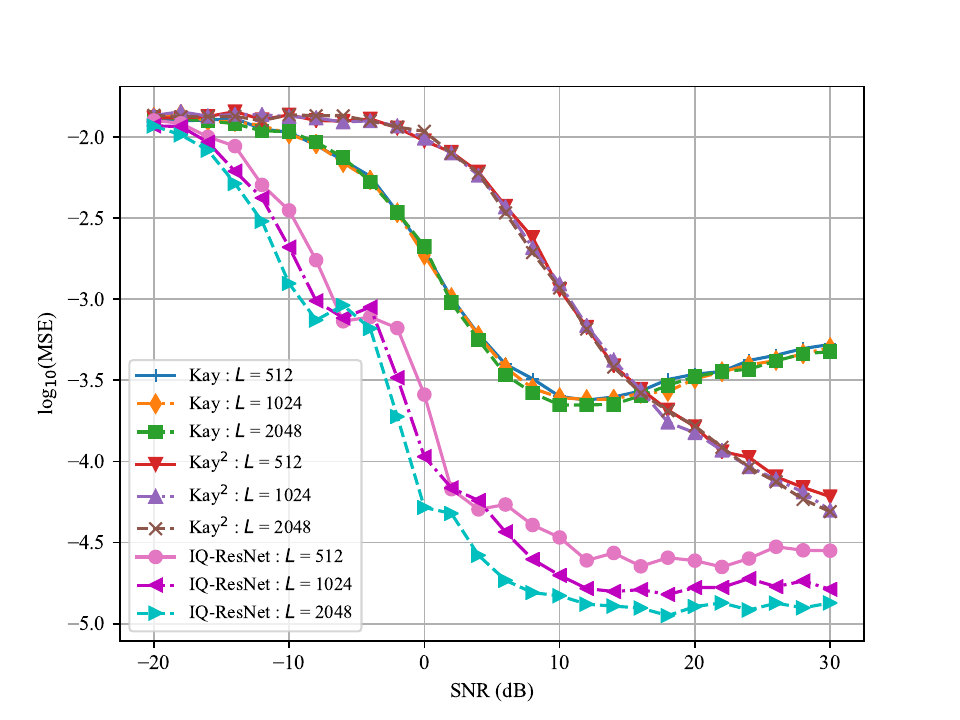}
\caption{\label{L}Effect of different sample lengths.}
\end{figure}

In the same vein, we consider how varying signal lengths impact the three methods. To this end, we generate datasets with signal lengths of 512, 1024, and 2048, each over-sampled at a rate of 8. The experimental results are represented in Fig. \ref{L}.
The results reveal that the Kay-based method and Kay$^2$-based method do not benefit from increased data length. 
In contrast, the IQ-ResNet method behaves differently. As the signal length increases from 512 to 2048, a noticeable improvement in the performance of the IQ-ResNet method is observed at different signal-to-noise ratios. This implies that as the signal length grows, a wider range of signal features becomes accessible. This expanded availability of signal characteristics empowers the IQ-ResNet method method to effectively utilize a more diverse set of distinctive signal features. Consequently, this phenomenon significantly contributes to the overall enhancement in CFO estimation.   


\subsubsection{Adaptability  Analysis}
\begin{figure}[t]
\centering
\includegraphics[width=0.4\textwidth]{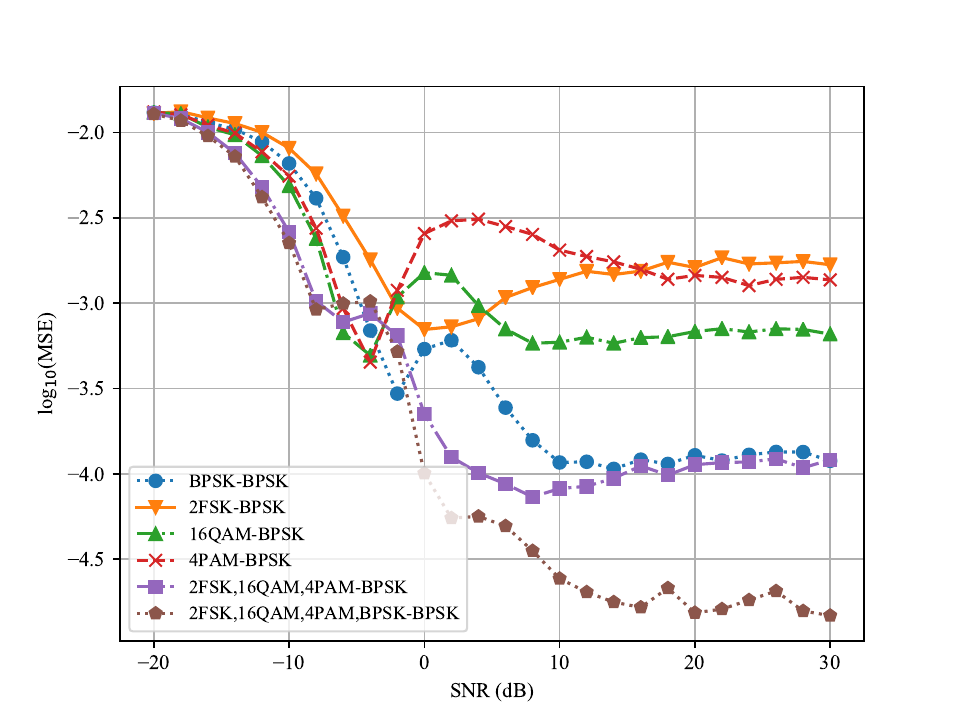}
\caption{\label{new}Performance of adaptability to new type of signals.}
\end{figure}
In order to validate the adaptability of our proposed IQ-ResNet method to new signals, we conduct experiments using different modulation types during training and evaluate its performance on a specific modulation type during testing. Specifically, the test set exclusively comprises samples with BPSK modulation, while the training set comprises four modulation types: BPSK, 2FSK, 16QAM, and 4PAM. The signal length is set to 1024, and the over-sampling rate is 8. Four models are trained with only one modulation which are referred to as BPSK-BPSK, 2FSK-BPSK, 16QAM-BPSK, and 4PAM-BPSK. One model is trained with all of the modulations which is referred to as and 2FSK,16QAM,4PAM,BPSK-BPSK. The last model is trained with all of the modulations but BPSK. We denote this model as 2FSK,16QAM,4PAM-BPSK. 

The results are shown in Fig. \ref{new}. Compared to using the same modulation type for prediction (BPSK-BPSK), the training models with another modulation (2FSK-BPSK and 16QAM-BPSK) fail to achieve superior predictions for the frequency offset value. However, when training samples involve multiple modulation types (2FSK, 16QAM and 4PAM), the performance of CFO estimation of BPSK signals increases and surpasses BPSK-BPSK even though BPSK has never been seen in the training process. 
As the variety of modulation types in the training samples expands, there is a progressive enhancement in the estimation performance. For example, the performance of 2FSK,16QAM,4PAM,BPSK-BPSK is better than that of 2FSK,16QAM,4PAM-BPSK at high SNR. This indicates that our proposed IQ-ResNet method enables the extraction of features from different modulation types for the purpose of estimating CFO of novel signal modulations, which further confirms the adaptability of our proposed method.



\section{Conclusion}
We have proposed a universal method for estimating CFO values which leverages IQ sequences and ResNet (denoted as the IQ-ResNet method). 
Our experiments have demonstrated the robustness of IQ-ResNet method across diverse scenarios. 
Notably, as the variety of modulation types present in the training set expands, there is a gradual enhancement in the estimation accuracy of IQ-ResNet for a specific modulation type in the test set  despite the modulation type is not the same. This phenomenon further validates the method's capability to deal with new kind of signals.

\ifCLASSOPTIONcaptionsoff
  \newpage
\fi

\bibliographystyle{IEEEtran}
\small
\bibliography{main}

\begin{thebibliography}{10}
\providecommand{\url}[1]{#1}
\csname url@samestyle\endcsname
\providecommand{\newblock}{\relax}
\providecommand{\bibinfo}[2]{#2}
\providecommand{\BIBentrySTDinterwordspacing}{\spaceskip=0pt\relax}
\providecommand{\BIBentryALTinterwordstretchfactor}{4}
\providecommand{\BIBentryALTinterwordspacing}{\spaceskip=\fontdimen2\font plus
\BIBentryALTinterwordstretchfactor\fontdimen3\font minus
  \fontdimen4\font\relax}
\providecommand{\BIBforeignlanguage}[2]{{%
\expandafter\ifx\csname l@#1\endcsname\relax
\typeout{** WARNING: IEEEtran.bst: No hyphenation pattern has been}%
\typeout{** loaded for the language `#1'. Using the pattern for}%
\typeout{** the default language instead.}%
\else
\language=\csname l@#1\endcsname
\fi
#2}}
\providecommand{\BIBdecl}{\relax}
\BIBdecl

\bibitem{2017Optimization}
X.~Jin, C.~Liu, and L.~Wen, ``Optimization of uav control system
  anti-jamming,'' \emph{Foreign Electronic Measurement Technology}, 2017.

\bibitem{2014A}
L.~Peng, J.~Sun, and X.~Wu, ``A joint rotational periodogram averaging and
  demodulation soft information carrier synchronization algorithm,''
  \emph{IEEE}, 2014.

\bibitem{2017A}
P.~Wang and L.~Kai, ``A non-data-aided frequency offset estimation algorithm of
  two co-frequency 16qam signals,'' in \emph{2016 International Conference on
  Audio, Language and Image Processing (ICALIP)}, 2017.

\bibitem{1055282}
D.~Rife and R.~Boorstyn, ``Single tone parameter estimation from discrete-time
  observations,'' \emph{IEEE Transactions on Information Theory}, vol.~20,
  no.~5, pp. 591--598, 1974.

\bibitem{7784355}
C.~Huang, J.~Suo, and T.~Liu, ``A novel autocorrelation approach for frequency
  estimation of a sinusoid,'' in \emph{2016 2nd International Conference on
  Control Science and Systems Engineering (ICCSSE)}, 2016, pp. 68--71.

\bibitem{9231178}
M.~Bertolucci, R.~Cassettari, and L.~Fanucci, ``Ccsds 131.2-b-1 frequency
  estimation trade-offs and a novel multi-algorithm fpga architecture,'' in
  \emph{2020 International Conference on Computing, Electronics \&
  Communications Engineering (iCCECE)}, 2020, pp. 224--229.

\bibitem{8649676}
H.~Abdzadeh-Ziabari, W.-P. Zhu, and M.~N.~S. Swamy, ``Timing and frequency
  synchronization and doubly selective channel estimation for ofdma uplink,''
  \emph{IEEE Transactions on Circuits and Systems II: Express Briefs}, vol.~67,
  no.~1, pp. 62--66, 2020.

\bibitem{8359649}
Y.~Wang and S.~Zhou, ``Non-data-aided frequency offset estimation for binary
  cpm signals,'' in \emph{2017 IEEE 17th International Conference on
  Communication Technology (ICCT)}, 2017, pp. 298--301.

\bibitem{2012Baseband}
T.~D. Chiueh, P.~Y. Tsai, and I.~Lai, \emph{Baseband Receiver Design for
  Wireless MIMO-OFDM Communications}.\hskip 1em plus 0.5em minus 0.4em\relax
  Baseband Receiver Design for Wireless MIMO-OFDM Communications, 2012.

\bibitem{937215}
E.~Rosnes and A.~Vahlin, ``Generalized kay estimator for the frequency of a
  single complex sinusoid,'' in \emph{ICC 2001. IEEE International Conference
  on Communications. Conference Record (Cat. No.01CH37240)}, vol.~10, 2001, pp.
  2958--2962 vol.10.

\bibitem{9245375}
B.~D. Bašić and M.~di~Buono, ``An analysis of early use of deep learning
  terms in natural language processing,'' in \emph{2020 43rd International
  Convention on Information, Communication and Electronic Technology (MIPRO)},
  2020, pp. 1125--1129.

\bibitem{2015Delving}
K.~He, X.~Zhang, S.~Ren, and J.~Sun, ``Delving deep into rectifiers: Surpassing
  human-level performance on imagenet classification,'' \emph{IEEE Computer
  Society}, 2015.

\bibitem{9935275}
T.~Chen, S.~Gao, S.~Zheng, S.~Yu, Q.~Xuan, C.~Lou, and X.~Yang, ``Emd and vmd
  empowered deep learning for radio modulation recognition,'' \emph{IEEE
  Transactions on Cognitive Communications and Networking}, vol.~9, no.~1, pp.
  43--57, 2023.

\bibitem{6415763}
A.~Vashist, R.~Chadha, M.~Kaplan, and K.~Moeltner, ``Detecting communication
  anomalies in tactical networks via graph learning,'' in \emph{MILCOM 2012 -
  2012 IEEE Military Communications Conference}, 2012, pp. 1--6.

\bibitem{8580824}
A.~Li, Y.~Me, S.~Xue, N.~Yi, and R.~Tafazolli, ``A carrier-frequency-offset
  resilient ofdma receiver designed through machine deep learning,'' in
  \emph{2018 IEEE 29th Annual International Symposium on Personal, Indoor and
  Mobile Radio Communications (PIMRC)}, 2018, pp. 1--6.

\bibitem{9992153}
M.~S. Chaudhari, S.~Majhi, and S.~Jain, ``Cnn-attention-dnn design for cfo
  estimation of non-pilot-assisted ofdm system,'' \emph{IEEE Communications
  Letters}, vol.~27, no.~2, pp. 551--555, 2023.

\bibitem{10042021}
S.~Zheng, X.~Zhou, L.~Zhang, P.~Qi, K.~Qiu, J.~Zhu, and X.~Yang, ``Toward
  next-generation signal intelligence: A hybrid knowledge and data-driven deep
  learning framework for radio signal classification,'' \emph{IEEE Transactions
  on Cognitive Communications and Networking}, vol.~9, no.~3, pp. 564--579,
  2023.

\end{thebibliography}

\end{document}